\documentclass[prb,twocolumn,superscriptaddress,floatfix]{revtex4-1}
\usepackage{epsfig}

\begin{document}
\title{Heterogeneous condensation of the Lennard-Jones vapor onto a nanoscale seed particle}

\author{Levent Inci}
\affiliation{Department of Chemistry, University of Saskatchewan, Saskatoon, SK, S7N 5C9, Canada}

\author{Richard K. Bowles}
\email{richard.bowles@usask.ca}
\affiliation{Department of Chemistry, University of Saskatchewan, Saskatoon, SK, S7N 5C9, Canada}

\date{\today}

\begin{abstract}

The heterogeneous condensation of a Lennard-Jones vapor onto a nanoscale seed particle is studied using molecular dynamics simulations. Measuring the nucleation rate and the height of the free energy barrier using the mean first passage time method shows that the presence of a weakly interacting seed has little effect on the work of forming very small cluster embryos but accelerates the rate by lowering the barrier for larger clusters. We suggest that this results from a competition between the energetic and entropic features of cluster formation in the bulk and at the heterogeneity. As the interaction is increased, the free energy of formation is reduced for all cluster sizes. We also develop a simple phenomenological model of film formation on a small seed that captures the general features of the nucleation process for small heterogeneities. A comparison of our simulation results with the model shows that heterogeneous classical nucleation theory provides a good estimate of the critical size of the film but significantly over-estimates the size of the barrier. 
\end{abstract}

\maketitle

\section{Introduction}

The heterogeneous condensation of a vapor onto a substrate is a key step in a wide range of chemical and physical processes that occur in both nature and technology. For example, dust and pollutant aerosol particles, ranging in size from several microns down to just a few nanometers, serve as cloud condensation nuclei in the atmosphere~\cite{kul04,kul08}, while nanoscale structured surfaces provide templates for the controlled nucleation and growth of complex materials such as colloidal metamaterials~\cite{zak98}. The general principles of heterogeneous nucleation on bulk surfaces are well known~\cite{debbook,kasbook}, but the details of how the process changes as the size of the heterogeneity becomes microscopic, and is the same size, or smaller, than the nanometer-sized fluctuations involved in the nucleation process itself, are not well understood. Recent experiments~\cite{win08} have shown that vapor condensation onto small clusters and nanoparticles, with diameters of $1-24$nm, occurs at supersaturations well below those predicted by classical nucleation theory~\cite{vol26,bec35,fre43,zeld43} (CNT) and the Kelvin relation, which suggests very small particles are far better at activating nucleation than expected. Similarly, Sear~\cite{sear06} found that a microscopic heterogeneity, consisting of a single fixed spin, was sufficient to accelerate the nucleation rate in a two-dimensional Ising model by as much as four orders of magnitude. This raises interesting questions regarding how small a heterogeneity can be and still have an influence on nucleation and whether there are practical limits to our ability to study homogeneous nucleation in the presence of small concentrations of impurities.  

The work of forming a critical size cluster is the central ingredient of CNT. The phenomenological expression for the work of forming an $n$-sized embryo of the new phase can be generally written as
\begin{equation}
\Delta G(n)=n\Delta\mu+an^{2/3}\mbox{, }\\
\label{eq:dgcnt}
\end{equation}
where $\Delta \mu$ is the difference in chemical potential between the bulk stable and metastable phases and $a$ is a geometric constant that is dependent on the shape of the embryo and proportional to the interfacial surface tension, $\gamma$.
The first term represents the free energy gained by moving particles from the metastable mother phase to the more stable phase and is negative, while the positive, second term accounts for the free energy cost of introducing an interface between the two phases. The functional form of Eq.~\ref{eq:dgcnt} highlights the activated nature of the nucleation process and shows that embryos must overcome a free energy barrier, $\Delta G(n^*)$, associated with making a critical cluster of size $n^*$, before they can grow spontaneously into the new phase. CNT then expresses the rate of nucleation as $J=A\exp[\Delta G(n^*)/kT]$, where the critical barrier height is obtained from  $\partial G(n)/\partial n=0$, $A$ is a kinetic prefactor that depends on the critical cluster size and the curvature at the top of the barrier, $k$ is Boltzmann's constant and $T$ is the temperature.

The simplest phenomenological model used in CNT to describe the condensation of a supersaturated vapor, which involves the capillarity assumptions of uniform bulk densities for both phases and a sharp interface characterized by the bulk, planar surface tension, gives the critical barrier and critical cluster size for the homogeneous nucleation as
\begin{equation}
\Delta G_{hom}(n^*)=\frac{16\pi}{3}\frac{v_l^2\gamma_{vl}^3}{(kT\ln S)^2}\mbox{,}\\
\label{eq:cgcnt}
\end{equation}
and
\begin{equation}
n^*=\frac{32\pi}{3}\frac{v_l^2\gamma_{vl}^3}{(kT\ln S)^3}\mbox{,}\\
\label{eq:cncnt}
\end{equation}
respectively. Here, $v_l$ is the volume per molecule in the bulk liquid phase, $S$ is the supersaturation and we have used $\Delta \mu\approx kT\ln S$ as well as assuming that the nucleus is spherical in shape.

In the presence of a macroscopic surface, nucleation can occur via a heterogeneous pathway. The nucleus then forms a droplet, contacting the wall with an angle, $\theta$, that characterizes the interaction of the material with the wall and is related to the wall-liquid ($\gamma_{wl}$) and wall-vapor ($\gamma_{wv}$) surface tensions through Young's equation, $\gamma_{vl} \cos\theta= \gamma_{wv}-\gamma_{wl}$. The extension of CNT to heterogeneous nucleation by Turnball~\cite{turn50} gives
\begin{equation}
\Delta G_{het}(n^*)=f(\theta)\Delta G_{hom}(n^*)\mbox{,}\\
\label{eq:ghet}
\end{equation}
where $0\le f(\theta)\le 1$ is solely a function of the contact angle. Eq~\ref{eq:ghet} shows that the barrier is always reduced because the wall contributes a portion of the interfacial free energy. In the limit where the contact angle goes to $\pi$, the liquid is non-wetting and $f(\theta)$ tends to one. Nucleation then occurs homogeneously in the bulk. If the liquid and surface are highly attractive so the liquid completely wets the wall, $\theta\rightarrow 0$ and $f(\theta)\rightarrow 0$, causing the barrier to go to zero. Despite the extremely simplified approach involved in CNT, recent simulation studies of heterogeneous nucleation in hard spheres colloids~\cite{auer03,cac04,cac05} and the ising model~\cite{wint09} have shown that Eq.~\ref{eq:ghet} is generally correct, as long as an additional term involving the three-phase contact line tension is included in $\Delta G(n)$. CNT also suggests that as the heterogeneity becomes more microscopic its ability to activate nucleation is greatly reduced~\cite{fle58}.

The goal of the present work is to quantify the effect of a very small seed particle on the condensation of the Lennard - Jones (LJ) vapor, by measuring both the nucleation rate and free energy barrier associated with heterogeneous nucleation, and compare the results with a simple thermodynamic model. Homogeneous nucleation in the LJ system has been studied by simulation~\cite{wold98,sen99,wed07,wed09}, as has heterogeneous nucleation on macroscopic surfaces~\cite{tox02}. However, studies involving small, seed particles in the LJ vapor have focused on examining the qualitative effects of highly attractive seeds, i.e. where the vapor - seed interaction is in the order of being ten times more attractive than the vapor-vapor interaction~\cite{suh08}. By studying the impact of seeds with interactions very similar to those of the vapor phase, we hope to understand how small heterogeneities perturb the bulk system. We can then directly address the question as to whether small impurities can seriously impact our ability to measure homogeneous nucleation rates, as well as assess the ability of CNT to predict heterogeneous nucleation on seed particles. The paper is organized as follows: Section 2 describes the development of a simple thermodynamic model that provides the appropriate CNT comparison for our nucleation rates and barriers calculated, using molecular dynamics (MD) simulations, in Section 3. Our discussion is contained in Section 4.

\section{Thermodynamic model}

\begin{figure}[h]
\includegraphics[width=3.5in]{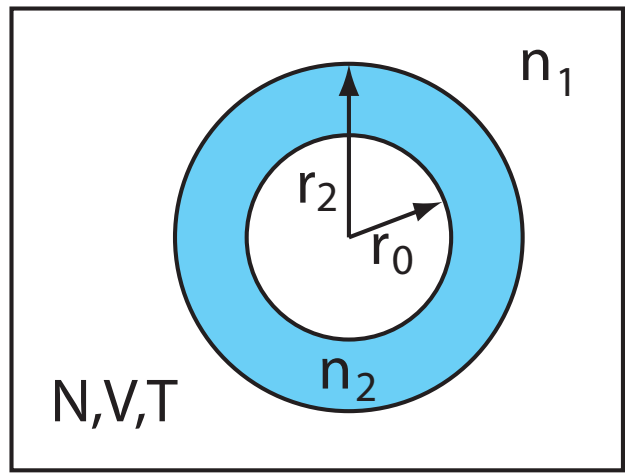}
\caption{$N,V,T$ model consisting of an insoluble heterogeneity of radius $r_0$ completely wet by a thin film of $n_2$ liquid atoms giving the film-seed composite a total radius of  $r_2$. The vapor is an ideal gas of $n_1$ atoms.}
\label{fig:model}
\end{figure}

We begin examining the problem of heterogeneous nucleation onto a nanoscale seed particle by considering the simple thermodynamic model described in Fig.~\ref{fig:model}, which represents the heterogeneous extension of the liquid drop model~\cite{wea93,reg03}. The system has a fixed total number of particles $N$, a fixed volume $V$, and a constant temperature, $T$. The heterogeneity is considered to be an insoluble, spherical, solid particle with radius $r_0$, while the liquid forms a uniform film of $n_2$ atoms that completely wets the particle giving rise to a film-particle composite of radius $r_2$ and leaving $n_1=N-n_2$ atoms in the vapor phase which is treated as an ideal gas. At constant $N,V,T$, the Helmholtz free energy, $F$, is the appropriate thermodynamic potential and variations in $F$ are given by
\begin{equation}
dF=dU-TdS\mbox{ ,}\\
\label{eq:df1}
\end{equation}
where, $U=U_1+U_2$, is the total internal energy and $S=S_1+S_2$ is the total entropy. Here, the subscripts $1$ and $2$ represent the vapor and liquid phases respectively. Variations in the $U$ are given by
\begin{equation}
dU_1=TdS-p_1dV_1+\mu_1dn_1\mbox{ ,}\\
\label{eq:du1}
\end{equation}
and
\begin{equation}
dU_2=TdS_2-p_2dV_2+\mu_2dn_2+\gamma_{12} dA_{12}+\gamma_{02}dA_{02}\mbox{ ,}\\
\label{eq:du2}
\end{equation}
where $p_1$ and $p_2$ are the respective pressures of each phase. We will assume that the interfaces between the phases are sharp, consistent with the capillarity approximation, so that the area of the liquid-vapor interface is given by, $A_{12}=4\pi r_2^2$, and the solid-liquid surface area, $A_{02}=4\pi r_0^2$.  $\gamma_{12}$ and $\gamma_{02}$ are the planar surface tensions of the interfaces respectively. Substituting Eqs.~\ref{eq:du1} and \ref{eq:du2} into Eq.~\ref{eq:df1} while using the conservation laws, $dV_1=-dV_2$ and $dn_1=-dn_2$, along with $dA_{12}=2dV_{2}/r_2$ and $dA_{01}=0$ yields,
\begin{equation}
dF=-(p_2-p_1-2\gamma_{12} /r_2)dV_2+(\mu_2-\mu_1)dn_2\mbox{ .}\\
\label{eq:df2}
\end{equation}

At equilibrium, $dF=0$, and Eq.~\ref{eq:df2} yields the expected equality of chemical potentials between the two phases and the Laplace relation for the difference in pressures across a curved interface,
\begin{equation}
\mu_2=\mu_1\mbox{ ,}\\
\label{eq:eqmu}
\end{equation}
\begin{equation}
\Delta p=p_2-p_1=2\gamma_{12} /r_2\mbox{ .}\\
\label{eq:eqp}
\end{equation}

A more detailed description of coexistence can be obtained by using the Gibbs-Duhem relation for each phase. For the vapor, we have $S_1dT-V_1dp_1+n_1d\mu_1=0$, which leads to
\begin{equation}
\mu_1(p_1)-\mu_1^{eq}(p_1^{eq})=kT\ln\frac{p_1}{p_1^{eq}}\mbox{ ,}\\
\label{eq:vgd}
\end{equation}
upon integration at constant $T$ and where we have chosen the reference pressure, $p_1^{eq}$, as the coexistence pressure of the vapor in contact with a liquid, with a planar interface. The Gibbs-Duhem relation for the liquid and its associated interface is, $S_2dT-V_2dp_2+n_2d\mu_2+A_{12}d\gamma_{12}=0$. If we assume that the surface tension is independent of pressure and that the liquid is incompressible, the integration of the Gibbs-Duhem relation gives,
\begin{equation}
\mu_2(p_2)-\mu_1^{eq}(p_1^{eq})=v_2(p_2-p_1^{eq})\mbox{ ,}\\
\label{eq:lgd}
\end{equation}
where $v_2$ is the volume per molecule in the liquid phase. We have also taken advantage of Eq.~\ref{eq:eqmu} and the fact that Eq.~\ref{eq:eqp} gives us $p^{eq}_2=p_1^{eq}$ in the limit $r_2\rightarrow\infty$, i.e. at a planar interface. Combining Eqs.~\ref{eq:eqp}, \ref{eq:vgd} and \ref{eq:lgd} yields the coexistence equation,
\begin{equation}
kT\ln\frac{p_1}{p_1^{eq}}=\frac{2\gamma_{12}}{r_2}v_2+v_2(p_1-p_1^{eq})\mbox{ ,}\\
\label{eq:kelvin}
\end{equation}
where
\begin{equation}
p_1=\frac{(N-n_2)kT}{V-(n_2v_2+(4/3)\pi r_0^3)}\mbox{ .}\\
\label{eq:p1}
\end{equation}

Eq.~\ref{eq:kelvin} is equivalent to the Kelvin relation and can be solved to find the equilibrium size of the film. In an open system, there would be one solution corresponding to the unstable equilibrium of the critical-sized film. Films that are thinner than the critical size tend to evaporate while thicker films grow spontaneously into a macroscopic drop as they consume molecules from a continually replenished vapor. However, in our closed system, $N$ is fixed and the growth of the liquid film depletes the number of molecules in the vapor phase, causing the supersaturation to decrease so that the film and vapor must eventually come into stable equilibrium.

\begin{figure}[h]
\includegraphics[width=3.5in]{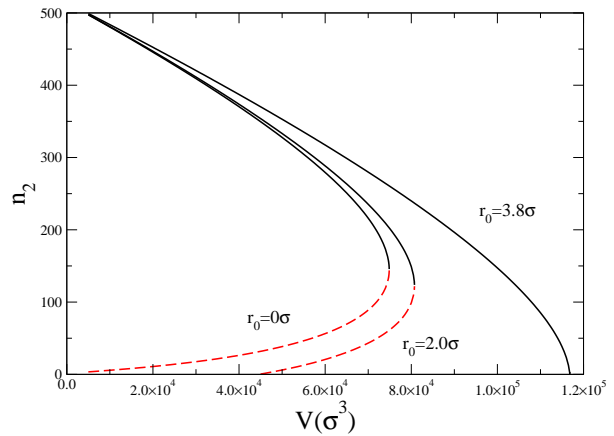}
\caption{Coexistence solutions for the film size, $n_2$, from Eq.~\ref{eq:kelvin}, with $N=512$ and $T^*=0.67$, in the presence of different size heterogeneous particles, $r_0/\sigma=0,2.0$ and $3.8$. The locally stable and unstable solutions are represented by the solid and dashed lines respectively.}
\label{fig:coexist}
\end{figure}

Fig~\ref{fig:coexist} shows the equilibrium film sizes obtained from Eq.~\ref{eq:kelvin} as a function of the total volume of the system for different sized seed particles, where we have used the LJ parameters of Baidakov et al~\cite{bai07} for $p_{eq},v_2$ and $\gamma_{12}$. All the systems exhibit a $V$, or evaporation volume, above which it is not possible to stabilize a film of any size and the stable equilibrium state consists of a dry particle surrounded by vapor. In the limit that $r_0\rightarrow 0$, we recover the solutions describing the formation of a liquid drop in the $N,V,T$ ensemble~\cite{reg03} and they have been included here for comparison with the heterogeneous nucleation case. The liquid drop model has two solutions below the evaporation volume. The large $n_2$ cluster belongs to the stable droplet in equilibrium with the vapor while the small droplet is the unstable, critical sized droplet that must be formed before the droplet can grow. We note two important features: firstly, there is always a barrier to the formation of the droplet, even at very small system volumes, which correspond to high initial supersaturations. Secondly, the droplet size remains finite at the evaporation volume. When a small insoluble particle is introduced into the system, we see that there are still two solutions at volumes just below the evaporation volume, but now the size of the small critical cluster tends to zero as the volume of the system is decreased so that there is no barrier to film formation at small enough volumes. As the size of the solid particle is increased, we eventually reach a seed size for which there is never a nucleation barrier associated with the formation of the film and the stable film grows continually from $n_2=0$ at the evaporation volume. This final result is consistent with the results of heterogeneous nucleation of films on bulk surfaces that exhibit complete wetting.

\begin{figure}[h]
\includegraphics[width=3.5in]{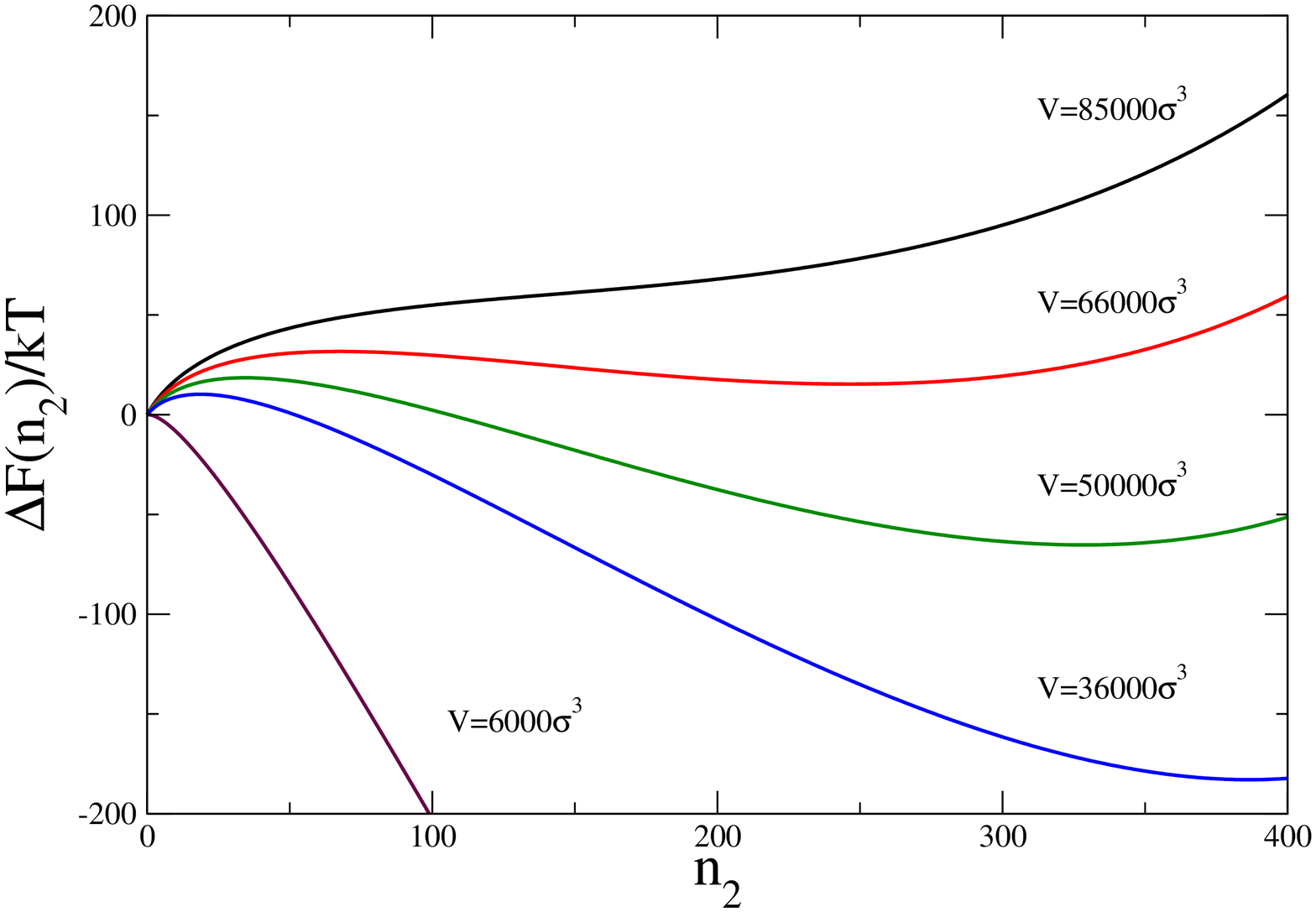}
\caption{$\Delta F(n_2)/kT$ as a function of $n_2$, with $N=512$, $r_0/\sigma=1.0$ and $T^*=0.67$, for different system volumes.}
\label{fig:fokt}
\end{figure}

Starting from Eq.~\ref{eq:df2}, using the expressions for the chemical potential from Eqs.~\ref{eq:vgd} and \ref{eq:lgd} and noting that $dV_2=v_2dn_2$ for a incompressible liquid, gives us an equation for $dF$ in terms of $dn_2$ that can be integrated at constant temperature to yield,
\begin{eqnarray}
\Delta F(n_2)&=&F(n_2)-F(0)=-n_2kT\ln\frac{p_1}{p_1^{eq}}\nonumber\\
&+&n_2(kT-v_2 p_1^{eq})+NkT\ln\frac{p_1}{p_0}\nonumber\\
&+&\gamma_{12}(A_{12}(n_2)-4\pi r_0^2)\mbox{ ,}\\
\label{eq:dff}
\end{eqnarray}
where $p_0=NkT/(V-(4/3)\pi r_0^3)$ is the pressure of the initial vapor before any film is formed. Eq.~\ref{eq:dff} represents the work of forming a film of $n_2$ molecules starting from an infinitely thin film with $n_2=0$ that wets the particle. In principle, there is also an additional term, $4\pi r_0^2(\gamma_{12}-\gamma_{01})$, which represents the free energy change associated with the wetting process, but this will be a negative constant and is not included here as it does not effect the probability of observing the critical sized cluster within the context of the model. Plots of Eq.~\ref{eq:dff} in Fig.~\ref{fig:fokt} for different volumes confirm that the small $n_2$ solutions represent unstable equilibrium solutions, while the large $n_2$ films are locally stable. They also highlight the global stability of the films and show that there is a range of volumes for which the thick film is locally stable, but metastable with respect to the dry heterogeneity.

\section{Molecular dynamics simulation studies}
Molecular dynamics simulations in the canonical, ($N,V,T$) ensemble are now used to study the condensation of a vapor onto a nanoscale heterogeneity. We model the composite system (vapor + heterogeneity) as a cut, but unshifted, Lennard - Jones mixture with the interaction potential,
\begin{equation}
V(r_{ij})=4\epsilon_{ij}\left[\left(\sigma_{ij}/r_{ij}\right)^{12}-\left(\sigma_{ij}/r_{ij}\right)^6\right]{ ,}\\
\label{pot}
\end{equation}
where $\epsilon_{ij}$ and  $\sigma_{ij}$ represent the energy and length interaction parameters. The vapor and heterogeneity are denoted as component 1and 2 respectively. We use $N=512$ vapor particles with $\epsilon_{11}=1.0$, and a single heterogeneity particle with interactions in the range $\epsilon_{12}=1.0-3.0$. This system size has been shown to be large enough that the depletion of the vapor phase caused by nucleation in the $N,V,T$ ensemble is small so the barriers are the same as those calculated in an open system~\cite{regsys}. All particles have the same size $(\sigma_{11}=\sigma_{22}=\sigma_{12}=$1$)$, the seed and gas particles have the same mass, $m$, and the potential is cut at $r_c=6.78\sigma_{ij}$, which should be long enough to approximate the full LJ potential and is consistent with a previous study of homogeneous condensation~\cite{wed09}. The simplicity of this model allows us to treat the heterogenous seed particle as another atom in the molecular dynamics simulation so it is free to translate throughout the system.

Simulations are carried out using the Gromacs Package~\cite{grom}, with the leap-frog integration scheme. The velocity rescaling thermostat is employed to maintain the system at a reduced temperature, $T^*=kT/\epsilon_{11}=0.67$, as this provides an efficient method for temperature control that does not appear to significant influence the kinetics of nucleation, even though the particle dynamics are perturbed in a non-physical way~\cite{thermo}. The volume of the simulation cell is chosen to ensure the initial starting conditions correspond to a particular supersaturation, defined as $S=p/p_{eq}$, where $p$ is the vapor pressure and $p_{eq}$ is the equilibrium coexistence pressure at $T$ as given by Ref.~\cite{bai07} Time is measure in reduced units, $t^*=t(\epsilon_{11}/m)^{1/2}/\sigma_{11}$ and we use periodic boundary conditions. 
 
The nucleation rate and free energy barrier are obtained using the mean first passage time (MFPT) approach introduced by Reguera et al~\cite{wed07b,wed07}. For each state point studied, we obtain 200 initial starting configurations by simulating the vapor phase in the absence of the heterogeneity for $10^6$ time steps, saving configurations every 5000 time steps. The heterogeneity is inserted randomly into the vapor, but ensuring that it is not placed within $\sigma$ of any vapor molecule. The MD trajectory is then followed as a function of time and the cluster size distribution is analyzed every 1000 time steps until the system is nucleated. Clusters are identified using the Frenkel cluster criteria~\cite{wold98}, which identifies liquid-like atoms as those particles that have at least five other atoms within a distance of $1.5\sigma$, and considers two liquid-like atoms within a distance of $1.5\sigma$ to be in the same liquid cluster. As a computational convieniece, we allow the seed particle to count as a neighbor when identifying liquid particles.

The MFPT for each cluster size, $\tau(n)$, is obtained by measuring the first time the largest cluster in the system reaches the size $n$ in a simulation trajectory, and averaging over the 200 trajectories. However, as a result of the intrinsic nature of the cluster dynamics and the fact that we only sample configurations periodically, the cluster growth is non-monotonic in time and does not proceed through a series of single particle additions or losses. It then becomes necessary to correct the time at which an $n$-sized cluster is observed for the first time in the simulation if a given cluster is missed in a trajectory all together, or if it is sampled out of order, i.e. when a small cluster is observed only after a larger cluster has already been sampled. We achieve this by assigning any small clusters that have not already been seen in a given trajectory, the same time that is assigned to the next largest cluster when it first appears~\cite{ivan}. 

When the barrier is high enough for the steepest descent approximations to hold, the MFPT can be represented by~\cite{wed07b},
\begin{equation}
\tau (n)= \frac {\tau_J}{2}[1+erf((n-n^*)c)]\mbox{,}\\
\label{eq:tau}
\end{equation}
where $n^*$ is the critical size, $\tau_J$ is the nucleation time, which is related to the steady state nucleation rate as $J=1/\tau_J V$, $c$ is associated with the Zeldovich factor, $Z$, as $c=\sqrt{\pi}Z$ and $erf(x)=2/\sqrt{\pi}\int_0^xe^{-x^2}dx$ is the error function. This allows us to extract $J,n^*$ and $c$ by using them as parameters in the fit of Eq.~\ref{eq:tau} to our data. The free energy barrier is given by~\cite{wed08},
\begin{equation}
\beta\Delta F(n)=\beta\Delta F(n_0)+\ln\left[\frac{B(n)}{B(n_0)}\right]-\int_{n_0}^{n}\frac{dn^{\prime}}{B(n^{\prime})}\mbox{,}\\
\label{eq:mfptf}
\end{equation}
where
\begin{equation}
B(n)=-\frac{1}{P_{st}(n)}\left[\int_n^b P_{st}(n^{\prime})dn^{\prime}-\frac{\tau(b)-\tau(n)}{\tau(b)}\right]\mbox{.}\\
\label{eq:bn}
\end{equation}
Here, $b$ is the upper absorbing boundary, taken as $b=60$ in this work, while $n_0=0$ represents the lower reflecting boundary and reference state. $P_{st}(n)$ is the steady state probability that the largest cluster in a given configuration from the ensemble of runs is of size $n$. 

\begin{figure}[h]
\includegraphics[width=3.5in]{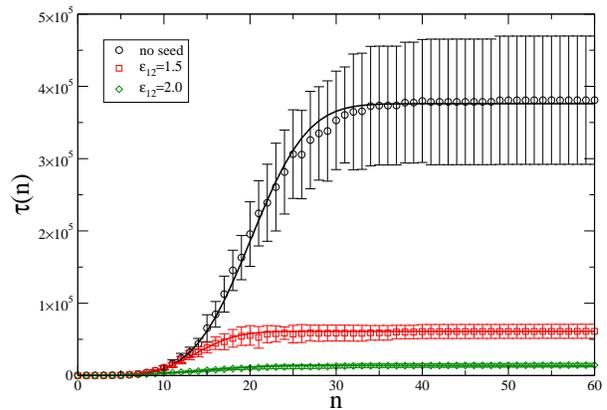}
\caption{$\tau(n)$ as a function of $n$, for systems with $N=512$, $S=10.43$ and a seed particle with $\epsilon_{12}=1.0$ (no seed), $1.5$ and $2.0$. The points represent data obtained from simulation and the error bars are the standard deviation of the block averages. The solid lines are best fits of Eq.~\ref{eq:tau} to the data.}
\label{fig:mfpt10}
\end{figure}

\begin{figure}[h]
\includegraphics[width=3.5in]{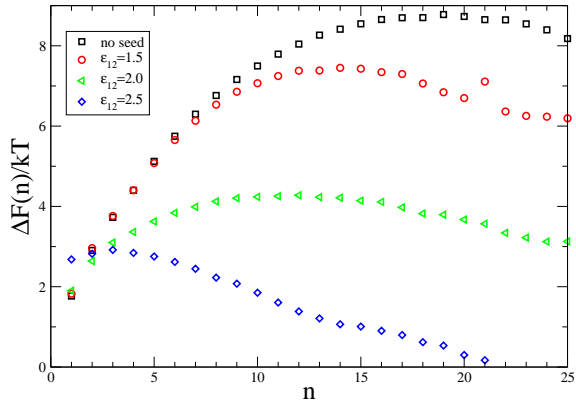}
\caption{$\Delta F(n)/kT$ as a function of $n$, for systems with $N=512$, $S=10.43$ and a seed particle with $\epsilon_{12}=1.0$ (no seed), $1.5,2.0$ and $2.5$.}
\label{fig:simfe10}
\end{figure}

A full summary of the nucleation rates, barrier heights and critical cluster sizes for all state points studied can be found in Table~\ref{tab:sim}.  Fig.~\ref{fig:mfpt10} shows $\tau(n)$ for systems with an initial vapor pressure of $S=10.43$, for different seed atoms with $\epsilon_{12}=1.0,1.5,2.0$ while Fig.~\ref{fig:simfe10} shows the free energy barriers. To estimate the error in our calculations, we divide the ensemble of 200 runs into 10 blocks and calculate $\tau(n)$ and $\Delta F(n)/kT$ for each block. The error bars in $\tau(n)$ represent the standard deviation of the block averages. The standard deviation in $\Delta F(n)/kT\approx0.5$ near $n^*$, but this grows to between 1-2 for larger cluster sizes. With $\epsilon_{12}=1.0$, the heterogeneity is simply an additional vapor molecule and we obtain $J=0.04\times10^{25} cm^{-3}s^{-1}$ and $n^*=20$ from the MFPT, where we have used the LJ argon parameters, $\sigma=0.3405$nm, $\epsilon/k =120$ K and $m=6.631\times10^{-26}$ kg to make the conversion from reduced units. The barrier calculations give $\Delta F(n)/kT=8.78$ and $n^*=19$. While our rate is approximately 33\% higher, and our barrier approximately $0.5kT$ lower, than that obtained by Wedekind et al~\cite{wed09}, the results are comparable when the slight difference in the definition of the supersaturation and the error in our calculations are taken into account. 

\begin{figure}[h]
\includegraphics[width=3.5in]{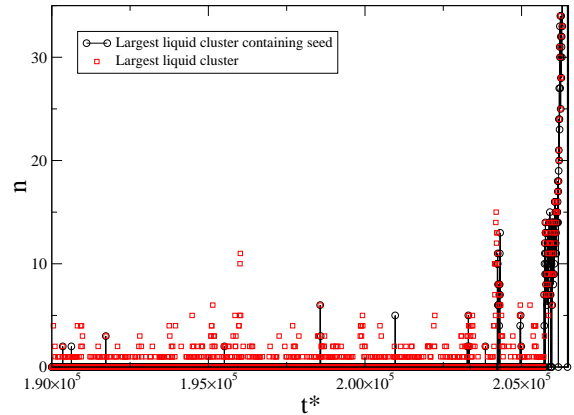}
\caption{Size of the largest cluster (squares), and size of the largest cluster containing the seed (joined circles)  as a function of time in a MD trajectory with $\epsilon_{12}=1.5$ and $S=10.43$}
\label{fig:traje15}
\end{figure}

The general effect of increasing the attraction of the seed particle is to increase the rate by lowering the barrier and decreasing the critical cluster size. With $\epsilon_{12}=1.5$, the work of forming small clusters with sizes 1-5 is the same as in the homogeneous case, with no seed, but then the heterogeneous barrier becomes lower for larger cluster sizes. To understand this, we analyze the MD trajectories and identify both the largest liquid cluster, which is our order parameter, and the largest cluster containing the seed. Fig.~\ref{fig:traje15} shows the two measures do not always coincide and most of the fluctuations involving small clusters do not contain the seed so the probability of seeing these clusters in the ensemble remains unchanged by the presence of the seed. However, the seed is always part of the cluster that eventually fluctuates over the barrier, suggesting that the added attraction is sufficient to help build the larger clusters and make them more probable that they would normally be, causing the barrier to decrease. 

\begin{table}
 \caption{Summary of simulation results. }
  \label{tab:sim}
  \begin{tabular}{ccccccc}
  \hline
$\epsilon_{12}$ & $S$ & $J\times10^{25} cm^{-3}s^{-1}$ & $n^*(Eq~\ref{eq:tau})$ &  $n^*(barrier)$ &  $\Delta F(n^*)/kT$  \\
\hline
1.0 &12.83 & 2.92 & 16 & 13 & 5.20 \\
1.0 &11.16 & 0.16 & 19 & 17 & 7.51 \\
1.0 &10.43 & 0.04 & 20 & 19 & 8.78 \\
1.5 &10.43 & 0.09 & 18 & 16 & 8.04 \\
2.0 &12.83 & 17.70 & 11 & 7 & 2.99 \\
2.0 &11.16 & 5.21 & 13 & 9 & 3.82 \\
2.0 &10.43 & 2.36 & 14 & 12 & 4.28 \\
2.0 &9.50 & 0.66 & 16 & 14 & 5.11 \\
2.0 &9.00 & 0.33 & 17 & 16 & 5.77 \\
2.0 &8.52 & 0.09 & 21 & 20 & 6.75 \\
2.0 &8.01 & 0.03 & 22 & 23 & 7.95 \\
\hline
\end{tabular}
\end{table}%

Some of the fluctuations in Fig.~\ref{fig:traje15} seem to suggest that the largest cluster liquid containing the seed intermittently drops to zero while the largest liquid cluster in the system grows. Following the system using the Stillinger cluster criteria~\cite{still63}, which counts particles within a distance of $1.5\sigma$ of each other as being part of the small clusters, allows us to identify those atoms that are loosely bound to the liquid-like particle, but may not have enough neighbors to be liquid-like themselves. We find the largest Stillinger cluster in the system and the largest Stillinger cluster containing the seed are identical for the larger clusters going over the barrier, but not when the small clusters are fluctuating in the metastable vapor. This emphasizes that the fluctuations involving small clusters are independent of the seed but that the seed is always connected to the largest liquid cluster as it grows, even if it is not always {\it wet} by five or more neighbors. 

\begin{figure}[h]
\includegraphics[width=3.5in]{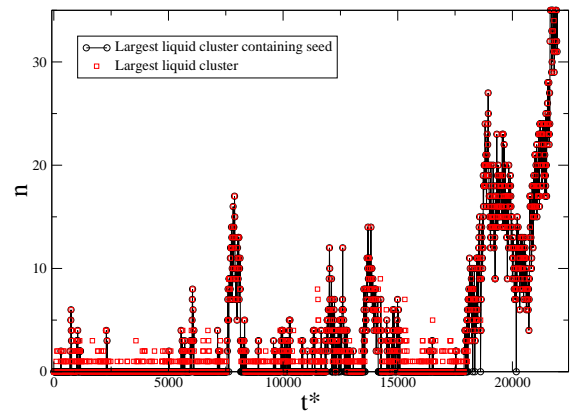}
\caption{Size of the largest cluster (squares), and size of the largest cluster containing the seed (joined circles)  as a function of time in a MD trajectory with $\epsilon_{12}=2.0$ and $S=10.43$}
\label{fig:traje2}
\end{figure}

The formation of monomers, dimers and trimers still appears to be relatively independent of the seed with $\epsilon_{12}=2.0$, (Fig.~\ref{fig:traje2}), but the seed appears in most, but not all, of the larger fluctuations, and this is reflected in the barrier. For example, at $S=10.43$, the seed is observed to be connected to trimers and 5-omer $55\%$ and $79\%$ of the time, respectively. The seed is found to be in clusters larger than $n=10$ more than $96\%$ of the time. As the supersaturation is decreased, the fluctuations of the bulk vapor naturally become rarer and most clusters are connected to the heterogeneity. Finally, when $\epsilon_{12}=2.5$, the nucleation times are very rapid and we are probably reaching the limits at which the method can be applied.

\begin{figure}[h]
\includegraphics[width=3.5in]{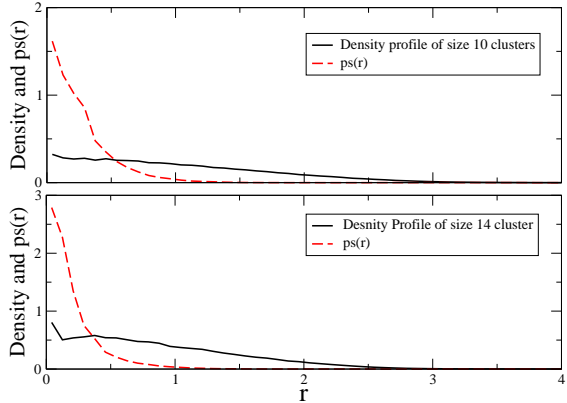}
\caption{The density profile (solid line) and $ps(r)$ (dashed line) for n=10 (top) and n=14 (bottom) clusters, with $S=10.43$ and $\epsilon_{12}=2.0$.}
\label{fig:psr}
\end{figure}

We can further quantify the location of the seed particle within the larger clusters by calculating $ps(r)$, the radial probability of finding the the seed a distance $r$ from the centre of mass of the cluster, for different $n$-size clusters, and comparing this with the density profile of the cluster (see Fig.~\ref{fig:psr}).  While the small cluster sizes of 10 and 14 give rise to broad density profiles where the density of the core is very much lower than the bulk, it is clear that the seed is generally located in the center of the cluster, so that it is effectively wet by the condensing vapor. 

\begin{figure}[h]
\includegraphics[width=3.5in]{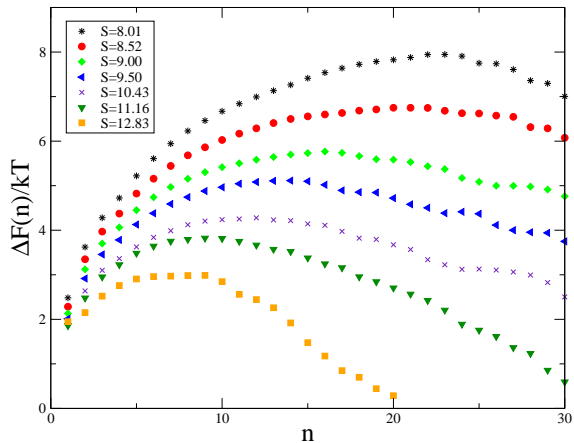}
\caption{$\Delta F(n)/kT$ as a function of $n$ at different supersaturations for a system with $N=512$ and a seed particle with $\epsilon_{12}=2.0$}
\label{fig:e2sim}
\end{figure}

\begin{figure}[h]
\includegraphics[width=3.5in]{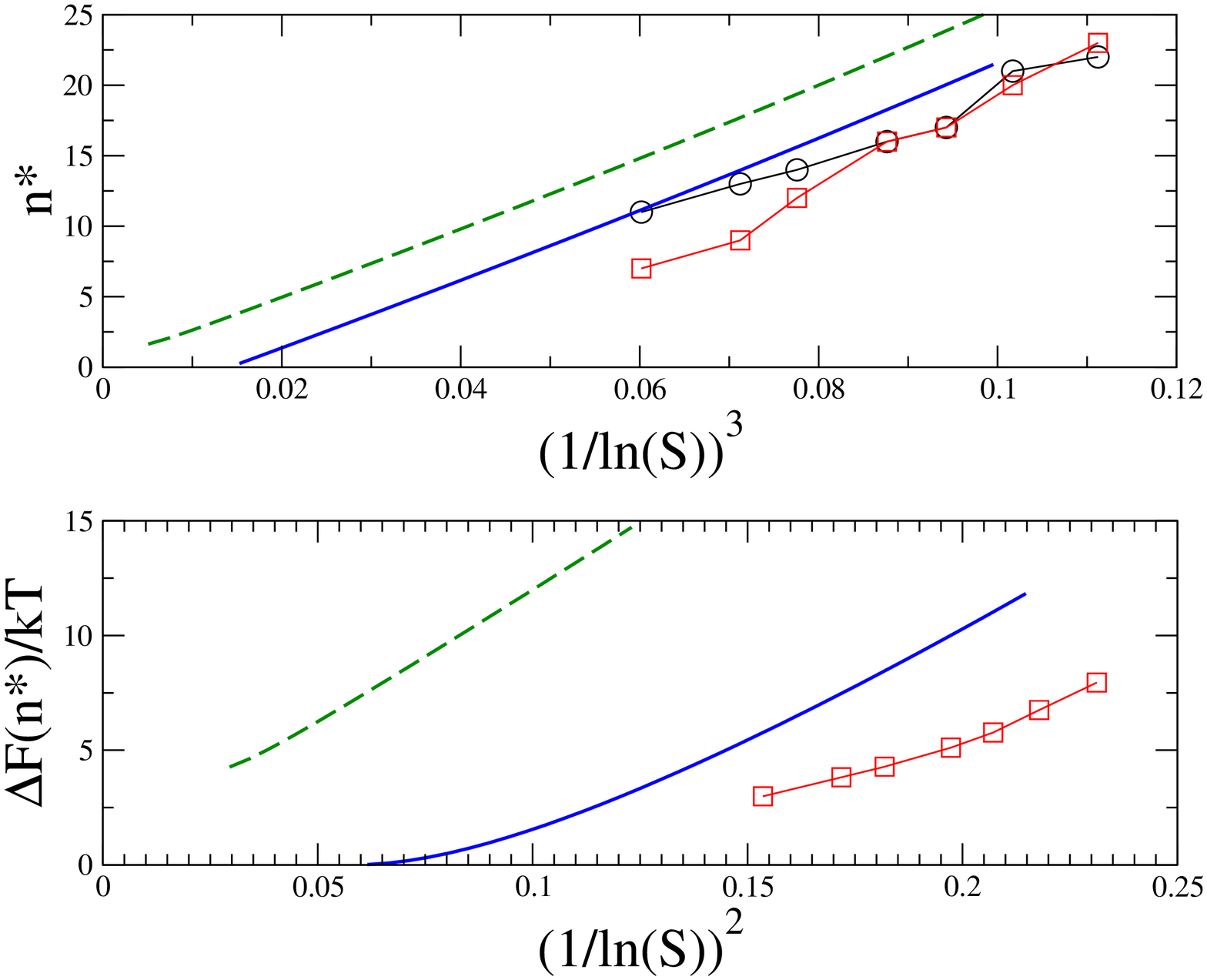}
\caption{(Top) Critical cluster size, $n^*$ as a function of $(1/\ln S)^3$. The circles and squares were obtained from fits of Eq.~ \ref{eq:tau} to our simulation values of $\tau(n)$ and the $\Delta F(n)/kT$ curves in Fig~\ref{fig:e2sim}, respectively. The solid line is obtained from the model using Eq.~\ref{eq:dff}, with $r_0/\sigma=1.0$, and the dashed line is the liquid drop model with no heterogeneity. (Bottom) Critical barrier height, $\Delta F(n^*)/kT$, as a function of $(1/\ln S)^2$. Symbols are the same as above.}
\label{fig:spinodal}
\end{figure}

To study the effects of supersaturation on heterogeneous nucleation, we calculate $\Delta F(n)/kT$ for a system containing a seed particle with $\epsilon_{12}=2.0$ over a range of $S$ (Fig.~\ref{fig:e2sim}) and compare the size of the critical nucleus and height of the barrier with our CNT-based model in Fig.~\ref{fig:spinodal}. The supersaturation for the model is defined as $p_0/p_{eq}$ in order to maintain the self-consistency. This assumes that the LJ vapor phase is well described by an ideal gas at the temperatures and densities studied, which is generally true away from the critical point~\cite{wed07,laas00}, but still leads to us underestimating the supersaturation in the model by approximately 14\% compared to the simulation at $S=10.43$.  Our results for heterogeneous nucleation essentially mirror those observed in homogeneous nucleation, in that the critical cluster size predicted by CNT is close to our simulation results, but the nucleation barriers are overestimated by a 100\% or more, which would lead to many orders of magnitude error in the rate.

\section{Discussion}
One of the goals of this work was to develop an understanding of how the presence of a small, nanoscale heterogeneity effects the condensation of a vapor and to quantify this by measuring both the nucleation rate and calculating the free energy barrier for the process. The obvious result is that the presence of the heterogeneity allows nucleation to occur faster, but when the attraction between the seed and the vapor is only marginally more attractive than the vapor-vapor interaction, we see the work of forming small clusters is essentially unchanged and the free energy is only lowered for the larger clusters. This appears to result from an interesting balance between the energetics of cluster formation and the entropic, or translational degrees of freedom, inherent in the system. There is a small energetic advantage to forming a monomer liquid particle that includes the seed, but there are many more ways of forming the monomer without the seed and these dominate the statistics in the density of states. Hence, there is no change in $\Delta F(n)$ for the small clusters. Larger clusters are rarer and the energetic advantage of including the seed in a cluster increases because it can have more neighbors. Eventually, the additional energy contribution dominates so the seed is connected to all the larger clusters and the barrier is lowered compared to homogeneous nucleation. 

We could completely decouple the heterogeneous and homogeneous processes by considering an order parameter that focuses just on $n$-sized clusters that include the seed. However, Fig.~\ref{fig:traje15} suggests the height of the nucleation barrier and the rate should remain unaffected because the seed is always part of the cluster that goes over the barrier and it is part of this cluster long before it reaches the critical size.  As a result, the MFPT of the larger clusters will not change and neither should the probability of finding the larger clusters within the ensemble of runs. The work of forming the smaller clusters would be expected to increase, as the probability of finding monomers and dimers etc decreases relative to the homogeneous case and the probability of finding no liquid-like atoms connected to the seed increases.

We also wished to make some quantitative assessment of the affect small heterogeneities may have on our ability to measure homogeneous nucleation. The presence of a single seed in the current simulations changes the rate of nucleation by one or two orders of magnitude, but it is important to note that we have relatively high concentration of heterogeneities in our system and we would expect, on the basis of the entropic arguments presented above, that the effect should decrease as the concentration of heterogeneities decreases.

Finally, the comparison of our simulation results with those of our model show that the simple capillarity base approach suffers the same degree of failure in heterogeneous nucleation as it does in the homogeneous case. The models are able to make relatively good predictions concerning the size of the critical nucleus but not the height of the barrier. Nevertheless, the model does capture the general features of heterogeneous nucleation, demonstrating both a spinodal limit in the supersaturation for a fixed sized heterogeneity and a critical size of activation, where the barrier to nucleation goes to zero for a completely wetting system above a certain size heterogeneity. There are also many corrections that could be introduced to the model to account for the small size of the system, such as size dependent surface tensions or the disjoining pressure, which provides a thermodynamic description of the interaction of the interfaces in a very thin film, that will probably lead to some improvement in the agreement. However, the very concepts of surface wetting and contact angle become poorly defined when describing processes involving heterogeneities that are of the small size considered here and more molecular approaches may be needed to describe nanoscale heterogeneous nucleation accurately.

\acknowledgments
The authors thank NSERC and CFI for financial support



\begin{thebibliography}{999}


\bibitem{kul04}M. Kulmala, H. Vehkam\"{a}ki, T. Pet\"{a}j\"{a}, M. Dal Maso, A. Lauri, V. M. Kerminen, W. Birmili and P. H. McMurry,  J. Aerosol Sci. {\bf 35}, 143 (2004).
\bibitem{kul08}  M. Kulmala and V. M. Kerminen,  Atmos. Res. {\bf 90}, 132, (2008).
\bibitem{zak98} A. A. Zakhidov, R. H.  Baughman, Z. Iqbal, C. Cui, I. Khayrullin, S. O. Dantas, J. Marti and V. G. Ralchenko,  Science {\bf  282}, 897 (1998).

\bibitem{debbook} P. G. Debenedetti,  {\it Metastable Liquids}; Princeton University Press: Princeton, 1996.
\bibitem{kasbook} D. Kashchiev, {\it Nucleation: Basic Theory With Applications}; Butterworth-
Heinemann, Oxford, 2000.

\bibitem{win08} P. M. Winkler, G. Steiner, A. Vrtala, H, Vehkamaki, M. Noppel, K. E. J. Lehtinen, G. P. Reischl, P. E. Wagner, M. Kulmala, Science {\bf 319}, 1374 (2008).

\bibitem{vol26} M. Volmer and A. Weber, Z. Phys. Chem. (Leipzig) {\bf 119}, 227 (1926).
\bibitem{bec35} R. Becker and W. D\"{o}ring, Ann. Phys. {\bf 24}, 719 (1935).
\bibitem{fre43} J. Frenkel, {\it Kinetic Theory of Liquids}; Clarendon, Oxford, 1946.
\bibitem{zeld43} Y. B. Zeldovich, Acta Physicochim. URSS {\bf 18}, 1 (1943).

\bibitem{sear06} R. P. Sear, J. Phys. Chem. B {\bf 110}, 4985 (2006).

\bibitem{turn50} D. Turnbull,  J. Chem. Phys. {\bf 18}, 198 (1950). 

\bibitem{auer03} S. Auer and D. Frenkel, Phys. Rev. Lett. {\bf 91}, 015703 (2003). 
\bibitem{cac04} A. Cacciuto, S. Auer and D. Frenkel, Nature {\bf 428}, 404 (2004). 
\bibitem{cac05} A. Cacciuto and D. Frenkel, Phys. Rev. E {\bf 72}, 041604 (2005).

\bibitem{wint09} D. Winter, P. Virnau and K. Binder, J. Phys.: Condens. Matter {\bf 21}, 464118 (2009). 

\bibitem{fle58} N. H. Fletcher, J. Chem. Phys. {\bf 29}, 572 (1958).

\bibitem{wold98} P. R. ten Wolde and D. Frenkel, J. Chem. Phys. {\bf 109}, 9901 (1998).
\bibitem{sen99} B. Senger, P. Schaaf, D. S. Corti, R. Bowles, D. Pointu, J.-C. Voegel and H. Reiss,  J. Chem. Phys. {\bf 110}, 6438 (1999).
\bibitem{wed07} J. Wedekind, J. Woelk, D. Reguera and R. Strey, J. Chem. Phys. {\bf 127}, 154515 (2007). 
\bibitem{regsys} J. Wedekind, D. Reguera and R. Strey, J. Chem. Phys. {\bf 125}, 214505 (2006).
\bibitem{wed09} J. Wedekind, G. Chkonia, J. W\"{o}lk, R. Strey and D. Reguera, J. Chem. Phys. {\bf 131}, 114506 (2009). 

\bibitem{tox02} S. Toxvaerd, J. Chem. Phys. {\bf 117}, 10303 (2002).

\bibitem{suh08} D. Suh,W. Yoon, M. Shibahara and S. Jung, J. Chem. Phys. {\bf 128}, 154523 (2008). 

\bibitem{wea93} C. L. Weakliem and H. Reiss, J. Chem. Phys. {\bf 99}, 9930 (1993).
\bibitem{reg03} D. Reguera, R. K. Bowles, Y. Djikaev, and H. Reiss, J. Chem. Phys. {\bf 118}, 340 (2003).

\bibitem{bai07} V. G. Baidakov, S. P. Protsenko, Z. R. Kozlova and G. G.Chernykh, J. Chem. Phys. {\bf  126}, 214505 (2007).  

\bibitem{grom} B. Hess, C.  Kutzner, D. van der Spoel and E. Lindahl,  J. Chem. Theory Comput. {\bf 4},  435 (2008).

\bibitem{thermo} J. Wedekind, D. Reguera, R. Strey, J. Chem. Phys.  {\bf 127}, 064501 (2007).


\bibitem{wed07b} J. Wedekind, R. Strey and D. Reguera, J. Chem. Phys. {\bf 126}, 134103 (2007). 
\bibitem{wed08} J. Wedekind and D. Reguera, J. Phys. Chem. B {\bf 112}, 11060 (2008).
\bibitem{ivan} S. E. M. Lundrigan and I. Saika-Voivod, J. Chem. Phys. {\bf 131}, 104503 (2009).

\bibitem{still63} F. H. Stillinger, J. Chem. Phys. {\bf 38}, 1486 (1963). 


 \bibitem{laas00} K. Laasonen, S. Wonczak, R. Strey and A.  Laaksonen,  J. Chem. Phys. {\bf 113}, 9741 (2000). 


\end{thebibliography}
\end{document}